\documentclass[12pt, draftclsnofoot, onecolumn]{IEEEtran}
\usepackage{amsmath}
\usepackage{amssymb}
\usepackage{amsthm}
\usepackage{cite}
\usepackage{color}
\usepackage{xcolor}

\usepackage{caption}
\usepackage{epsfig,latexsym}
\usepackage{float}
\usepackage{fancyhdr}
\usepackage{graphicx}
\usepackage{indentfirst}
\usepackage{mdwmath}
\usepackage{mdwtab}
\usepackage{subfigure}
\usepackage{setspace}
\usepackage{times}
\usepackage{url}
\usepackage{multirow}

\newtheorem{remark}{Remark}
\newtheorem{theorem}{Theorem}

\newtheorem{lemma}{Lemma}

\newtheorem{corollary}{Corollary}

\newtheorem{proposition}{Proposition}

\begin{document}

\title{Non-Orthogonal Multiple Access in UAV-to-Everything (U2X) Networks}
\author{Tianwei Hou,~\IEEEmembership{Student Member,~IEEE,}
        Yuanwei Liu,~\IEEEmembership{Senior Member,~IEEE,} \\
        Zhengyu Song,
        Xin Sun,
        and Yue Chen,~\IEEEmembership{Senior Member,~IEEE,}

\thanks{This work is supported by the Fundamental Research Funds for the Central Universities under Grant 2016RC055. Part of this work was submitted to the IEEE Global Communication Conference, Waikoloa, USA, Dec. 2019~\cite{Hou_UAV_3D_Glob}.}
\thanks{T. Hou, Z. Song and X. Sun are with the School of Electronic and Information Engineering, Beijing Jiaotong University, Beijing 100044, China (email: 16111019@bjtu.edu.cn, songzy@bjtu.edu.cn, xsun@bjtu.edu.cn).}
\thanks{Y. Liu and Yue Chen are with School of Electronic Engineering and Computer Science, Queen Mary University of London, London E1 4NS, U.K. (e-mail: yuanwei.liu@qmul.ac.uk, yue.chen@qmul.ac.uk).}
}

\maketitle

\begin{abstract}
This article investigates the non-orthogonal multiple access (NOMA) enhanced unmanned aerial vehicle (UAV)-to-Everything (U2X) frameworks. A novel 3-Dimension framework for providing wireless services to randomly roaming NOMA receivers (Rxs) in the sphere space is proposed by utilizing stochastic geometry tools. In an effort to evaluate the performance of the proposed framework, we first derive closed-form expressions for the outage probability and the ergodic rate of paired NOMA Rxs. For obtaining more insights, we investigate the diversity order and the high signal-to-noise (SNR) slope of NOMA enhanced U2X frameworks. We also derive the spectrum efficiency in both NOMA and orthogonal multiple access (OMA) enhanced U2X frameworks. Our analytical results demonstrate that the diversity order and the high SNR slope of the proposed framework are $m$ and one, respectively. Numerical results are provided to confirm that: i) the proposed NOMA enhanced U2X frameworks have superior outage performance and spectrum efficiency compared with the OMA-enhanced U2X frameworks; and ii) for the case of fixed LoS probability, the outage performance of paired NOMA Rxs mainly depends on users with poor channel conditions.

\end{abstract}

\begin{IEEEkeywords}
Non-orthogonal multiple access, stochastic geometry, unmanned aerial vehicles, UAV-to-Everything.
\end{IEEEkeywords}

\section{Introduction}

Unmanned aerial vehicle (UAV) communications, standing as a supplementary communication scenario in the next generation communication systems and beyond, have received considerable attention in recent years~\cite{UAV_general_Liu}, such as industry standardization third Generation Partnership Project Long-Term Evolution Advanced (3GPP-LTE-A) standard, the fifth generation (5G) New Radio standard, and the next general digital TV standard (ATSC 3.0)~\cite{Liu_NOMA_beyond}.
UAV communication is capable of providing access services where there is a temporary need for network resources, i.e., during temporary events and after disasters in the remote areas for Air-to-Ground (A2G) communications and Air-to-Air (A2A) communications~\cite{UAV_civil_use,UAV_Chandre_Mag}.
UAV communications differ significantly from conventional ground base station (BS) communications in terms of the mobility, energy constraint, user distributions, as well as the large-scale and small-scale propagations~\cite{UAV_general_Liu,UAV_Channel}. Compared to conventional BS communications, UAV communication is capable of offering stronger received power to users because of the existence of line-of-sight (LoS) propagation between UAV and users. These advantages have stimulated interest in the design of UAV communication protocols for effectively utilizing the resources of UAV networks.

With significant advancements in multiple technologies, non-orthogonal multiple access (NOMA) has been recently recognized as a promising solution to realize the performance requirements of next-generation mobile networks and beyond, i.e., enhanced Mobile Broadband (eMBB) and massive Machine Type Communications (mMTC)~\cite{Islam_NOMA_survey,Dai_NOMA_survey,Shin_NOMA_cellular,NOMA_mag_DingLiu}. More specifically, in contrast to the conventional orthogonal multiple access (OMA) techniques, NOMA is capable of exploiting the available resources more efficiently by providing enhanced spectrum efficiency and massive connectivity on the specific channel conditions of users~\cite{PairingDING2016}, and it is capable of serving multiple users at different quality-of-service (QoS) requirements in the same resource block for both eMBB and MTC networks~\cite{NOMA_5G_beyond_Liu,Islam_NOMA_survey}. To be more clear, NOMA technique sends the signal to multiple users simultaneously by power domain multiplexing within the same frequency, time and code block. The basic principles of NOMA techniques rely on the employment of superposition coding (SC) at the transmitter (Tx) and successive interference cancelation (SIC) techniques at the receiver (Rx)~\cite{heterNOMA_Qin}, and hence multiple accessed Rxs can be realized in the power domain via different power levels for Rxs in the same resource block.

\subsection{Motivations and Prior Work}

Previous research in A2G networks~\cite{UAV_finite_downlink,Rayleigh_UAV} mainly considered that multiple terrestrial Rxs were located in a disc on the ground, whereas A2A networks~\cite{UAV_Chandre_Mag,Saad_D2D_UAV} mainly considered that multiple UAVs were located in a disc in the sky with the same height.
The distinctive channel propagations for both A2G and A2A networks were investigated in~\cite{UAV_Channel}, where different types of small-scale fading channels were summarized to demonstrate the significant differences for channel propagation between UAV communications and conventional BS communications. It is demonstrated that both horizontal distance and vertical distance between UAV and users affect the small-scale fading channels. A downlink A2G network was proposed in~\cite{UAV_finite_downlink}, where multiple UAVs are distributed in a finite 3-D network with Nakagami-$m$ fading channels. An uniform binomial point process was invoked to model the finite 3D networks. It is also worth noting that Rayleigh fading channel~\cite{Rayleigh_UAV}, which is a well-known model in scattering environment, can be also used to model the UAV channel characteristics in the case of large elevation angles in the mixed–-urban environment. Generally speaking, Nakagami-$m$ distribution and Rician distribution are used to approximate the fluctuations in fading channels with LoS propagations. It is also worth noting that the fading parameter of Nakagami-$m$ fading $m=\frac{(K+1)^2}{2K+1}$, the distribution of Nakagami-$m$ is approximately Rician fading with parameter $K$~\cite[eq. (3.38)]{wireless_communication_goldsmith}.
Recently, a new probability of LoS scenario was proposed for A2G communications~\cite{3GPP_36.777}, where the existence of LoS propagation is based on the height of the UAV, the horizontal distance between the UAV and users, the carrier frequency and type of environment. For instance, a trajectory design and a power control strategy for multi-UAV networks, which were based on position data collected from Twitter, has been proposed in~\cite{Liuxiao_Trajectory_multi-UAV}. The simulation results demonstrated that throughput gains of about 17 percent were achieved by applying a Q-learning approach. A UAV assisted cooperative jamming for physical layer security was proposed in~\cite{cooperative_jamming_UAV}, where UAV can optimize its trajectory for jamming eavesdroppers.
Two possible paradigms for UAV assisted cellular communications were proposed in~\cite{UAV_Trajectory_shuowen}, namely, cellular-enabled UAV communication and UAV-assisted cellular communication. A multiple-input multiple-output (MIMO) assisted UAV network was proposed in~\cite{UAV_multibeam_liangliu}, where the UAV serves multiple users through multi-beam simultaneously. Since the number of connected devices may be practically large, which is in need of massive connectivity, new research on UAV under emerging next generation network architectures, i.e., NOMA, is needed.

Integrating NOMA into UAV-to-Everything (U2X) networks is considered to be a promising technique to significantly enhance the spectrum efficiency and energy efficiency for UAV communications in the next generation wireless system and beyond, where UAVs are deployed with multi-antenna to serve ground users by NOMA~\cite{Hou_Single_UAV}. A general introduction of NOMA enhanced UAV communications has been proposed in~\cite{UAV_general_Liu}. Three case studies, i.e., performance evaluation, joint trajectory design, and machine learning enhanced UAV deployment, were carried out in order to better understand NOMA enabled UAV networks.
In UAV-enabled wireless communications, the total UAV energy is limited, which includes propulsion energy and communication related energy~\cite{energy_consumption_UAV}. Therefore, integrating UAVs and NOMA into cellular networks is considered to be a promising technique to significantly enhance the performance of terrestrial users in the next generation wireless system and beyond, where the energy efficiency and spectrum efficiency can be greatly enhanced in downlink transmission to minimize communication related energy~\cite{Cellular-UAV_mag}.
A cooperative UAV network was proposed in~\cite{UAV_relay}, where multiple UAVs, which are distributed in a 2D disc located in the sky, are used as a flying relay in NOMA assisted wireless backhaul A2A network. A NOMA enhanced multi-UAV network was proposed in~\cite{Hou_multi_UAV}, where the imperfect SIC scenario is taken into account in the large-scale cellular A2G networks in order to provide more engineering insights. All the flying UAVs and terrestrial users are located according to 2D HPPPs.
A NOMA assisted uplink scenario of UAV communication was proposed in~\cite{Uplink_NOMA_UAV}, where two special cases, i.e., egoistic and altruistic transmission strategies of the UAV, were considered to derive the optimized solutions. A UAV assisted millimeter-wave air-to-everything networks was proposed in~\cite{Han_millimeter_UAV}, where aerial access points provide access services to users located on the ground, air, and tower. The buildings were modeled as a Boolean line-segment process with the fixed height.
The outage performance of NOMA downlink transmission in Nakagami-$m$ fading channels was evaluated in~\cite{Nakagami_Hou}, which indicates that NOMA sacrifices the outage performance of the user with poorer channel gain while increasing the outage performance of the user with better channel gain dramatically.
The trajectory of movable UAV in both OMA and NOMA scenario was designed in~\cite{Mobility_UAV_Trajectory}, where new algorithms were proposed to maximize the average rate of ground users.

The previous contributions~\cite{Hou_Single_UAV,UAV_relay,Hou_multi_UAV,Uplink_NOMA_UAV,Han_millimeter_UAV} mainly focus on NOMA in A2G networks, where multiple Rxs are distributed in the 2D plane.
Thus, the performance of NOMA enhanced U2X networks, where Rxs are distributed in a 3D sphere space, is still in its infancy.
To-date, to the best of our knowledge, there has been no existing work intelligently investigating the performance of NOMA enhanced U2X frameworks, particularly with the focus of 3-D distributed Rxs, which motivates us to develop this treatise.
In this article, inspired by the ad hoc networks and D2D networks~\cite{Yi_D2D}, a U2X framework for intelligently investigating the effect of NOMA enhanced U2X framework performance is desired. The motivation of proposing U2X frameworks is that the U2X framework can be deployed for multiple purposes properly in the next generation wireless systems and beyond, i.e., A2G networks, A2A networks, Air-to-Vehicular (A2V) networks. In this article, we will develop the first comprehensive model aimed at the downlink analysis of a finite U2X framework using tools from stochastic geometry, which is capable to provide the mathematical paradigm to model the spatial randomness of 3D sphere U2X frameworks.

\subsection{Contributions}

This paper focuses on the application of NOMA enhanced 3D sphere U2X frameworks, which is also applicable for A2G, A2A, and A2V communications. Based on the proposed framework, the primary theoretical contributions can be summarized as follows:

\begin{itemize}
  \item We propose a novel NOMA enhanced U2X framework, where stochastic geometry approaches are invoked to model the 3D sphere distributions of Rxs. By utilizing this framework, both LoS and NLoS links are considered to illustrate the general case of NOMA enhanced U2X frameworks.
  \item We derive closed-form expressions in terms of outage probability for paired NOMA Rxs in the proposed framework. Both exact results and asymptotic results are derived for obtaining engineering insights. Furthermore, diversity orders are obtained for the paired NOMA Rxs based on the developed outage probability. The obtained results confirm that the diversity order of the proposed framework is determined by the fading parameters $m$.
  \item We derive closed-form expressions in terms of ergodic rate for paired NOMA enhanced Rxs. We obtain high SNR slopes for the paired NOMA Rxs based on the developed ergodic rate. The obtained results confirm that the high SNR slopes of NOMA enhanced U2X framework is one for both LoS and NLoS scenarios.
  \item We also derive closed-form expressions for OMA scenario in terms of outage probability and ergodic rate. We show that the NOMA enhanced U2X framework has superior performance over OMA enhanced U2X framework. Our Analytical results demonstrate that for the probability of LoS scenario, the outage performance of paired NOMA users mainly depends on the NLoS scenario.
\end{itemize}

\subsection{Organization and Notations}

The rest of the paper is organized as follows. In Section \uppercase\expandafter{\romannumeral2}, a model of U2X transmission framework is investigated in wireless networks, where NOMA technique is invoked. Analytical results are presented in Section \uppercase\expandafter{\romannumeral3} to show the performance of NOMA enhanced U2X frameworks. Our numerical results are demonstrated in Section \uppercase\expandafter{\romannumeral4} for verifying our analysis, which is followed by the conclusion in Section \uppercase\expandafter{\romannumeral5}.

\section{System Model}

\begin{figure*}[t!]
\centering
\includegraphics[width =4in]{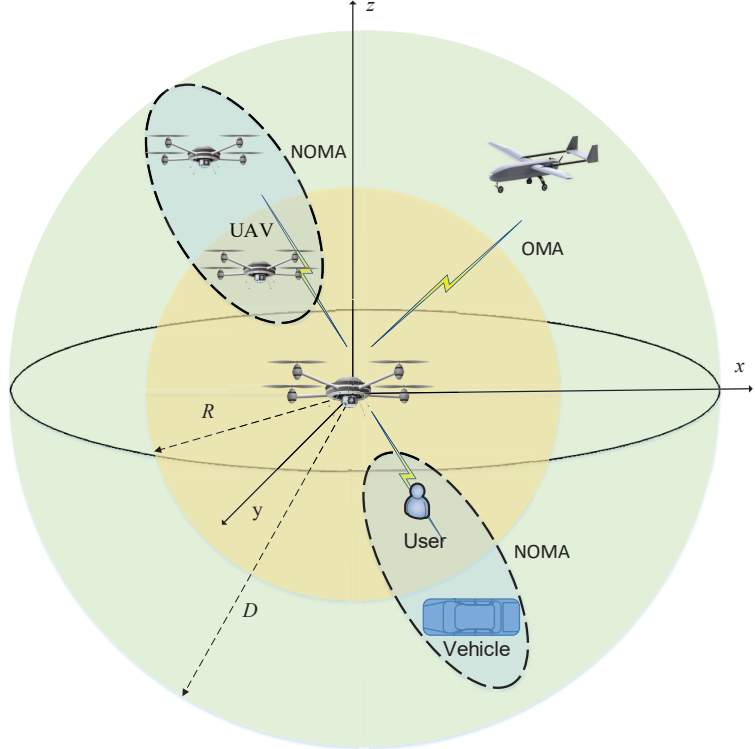}
\caption{Illustration of a typical U2X framework supported by omni-antenna.}
\label{system_model}
\end{figure*}

Consider a NOMA enhanced U2X downlink communication scenario in which a UAV equipped with a single omni transmitting antenna is communicating with multiple Rxs equipped with a single omni transmitting antenna each. Fig.~\ref{system_model} illustrates the wireless communication model with a single UAV.

\subsection{System Description}

For tractability purpose, the UAV cell coverage space is a sphere, denoted by $\mathcal{V}^3$. The radius of the sphere is $D$, and the Tx-UAV is located at the center of $\mathcal{V}^3$. It is assumed that the near Rxs and far Rxs are uniformly distributed according to homogeneous poisson point process (HPPP), which is denoted by $\Psi$ and associated with the density $\lambda$, within small sphere $\mathcal{V}_1^3$ and large hollow sphere $\mathcal{V}_2^3$ with radius $R$ and $D$ ($D > R$), respectively. For simplicity, we only focus our attention on investigating a typical Rx pairing in this treatise, where two Rxs, $w$-th Rx and $v$-th Rx, are grouped to deploy NOMA transmission protocol.

\subsection{Channel Model}

Consider the use of a composite channel model with two parts, large-scale fading and small-scale fading. $L$ denotes the large-scale fading, which represents the path loss between the Tx-UAV and Rx. It is assumed that large-scale fading and small-scale fading are independent and identically distributed (i.i.d.). Generally speaking, the large-scale fading between the Tx-UAV and Rxs can be expressed as

\begin{equation}\label{large scale fading,eq1}
L({d}) = \left\{ \begin{array}{l}
d^{ - \alpha },{\kern 1pt} \emph{if} {\kern 5pt} {d} > {r_0}\\
r_0^{ - \alpha }, \emph{otherwise}
\end{array} \right.,
\end{equation}
where $\alpha$ denotes the path loss exponent, and the parameter $r_0$ avoids a singularity when the distance is small. For simplicity, it is assumed that the radius of small sphere is grater than $r_0$, i.e., $R>r_o$.

In order to better illustrate the LoS propagation between the Tx-UAV and Rxs, the probability density functions (PDFs) of small-scale fading is defined by Nakagami-\emph{m} fading as
\begin{equation}\label{channel matrix,eq3}
{f}(x) = \frac{{m^m {x^{{m} - 1}}}}{{\Gamma ({m})}}{e^{ - {{mx}}}},
\end{equation}
where $m$ denotes the fading parameter, and $\Gamma ({m})$ denotes Gamma function. Note that $\Gamma ({m})=(m-1)!$ when $m$ is an integer. For notation simplicity, $h_{w}$ and $h_{v}$ denote the small-scale channel coefficients for the near Rx and the far Rx, respectively.

Thus, the received power for the $w$-th Rx from the Tx-UAV is given by
\begin{equation}\label{received user power}
{P_{w}} = {P_u}{L_w}{\left| {{{h}_{w}}} \right|^2},
\end{equation}
where $P_u$ denotes the transmit power of the Tx-UAV. Besides, in practical wireless communication systems, obtaining the channel state information (CSI) at the transmitter or receiver is not a trivial problem, which requires the classic pilot-based training process. Therefore, in order to provide more engineering insights, it is assumed that the CSI of UAVs is partly known at the typical user, where only distance information between UAVs and typical user is required.

\section{Performance Evaluations}

In this section, we discuss the performance of downlink NOMA enhanced U2X frameworks. In this paper, fixed power allocation is employed at the Tx-UAV. New channel statistics, outage probabilities, ergodic rates, and spectrum efficiency are illustrated in the following four subsections.

\subsection{New Channel Statistics}

In this subsection, we derive new channel statistics for NOMA enhanced U2X frameworks, which will be used for evaluating the outage probabilities and ergodic rates in the following subsections.

\begin{lemma}\label{lemma1:new channel state}
Assuming that Rxs are located according to HPPP in the space of Fig.~\ref{system_model}. Therefore, the Rxs are independently and identically distributed in the coverage space, and the PDFs of far Rxs and near Rxs are given by

\begin{equation}\label{PDF of m'}
{f_{v}}\left( x \right) = \frac{3r^2}{{ ( D^3 - R^3)}}, {\rm{if}}~~ R<r<D,
\end{equation}
and
\begin{equation}\label{PDF of m}
{f_w}\left( x \right) = \frac{3r^2}{{ (R^3-r_0^3)}}, {\rm{if}}~~  r_0<r<R,
\end{equation}
respectively.
\begin{proof}
According to HPPP, the PDF of the far Rxs can be given by
\begin{equation}\label{PDF of m'}
{f_{v}}\left( x \right) = \frac{{{\lambda}{\Psi}4\pi r^2}}{{{\lambda}{{\Psi}}\left( {\frac{4}{3}  \pi D^3 - \frac{4}{3} \pi R^3} \right)}}.
\end{equation}
After some algebraic handling, Lemma 1 is proved.
\end{proof}
\end{lemma}

\subsection{Outage Probabilities}

In this subsection, we first focus on the outage behavior of far Rx $v$, who is the Rx with poorer channel gain. The fixed power allocation strategy is deployed at the Tx-UAV, which the power allocation factors $\alpha_w^2$ and $\alpha_{v}^2$ are constant during transmission. It is assumed that the target rate of the near Rx and far Rx are $R_{w}$ and $R_{v}$, respectively. Therefore, the outage probability of the $v$-th Rx is given by

\begin{equation}\label{Outage}
{\rm P}_{v} = \mathbb{E} \left( {\log \left( {1 + \frac{{P_u {{\left| {{h_{v}}} \right|}^2}d_v^{-\alpha} \alpha _{{v}}^2}}{{\sigma^2 + P_u {{\left| {{h_{w}}} \right|}^2}d_v^{-\alpha}\alpha_w^2  }}} \right) < {R_{v}}} \right),
\end{equation}
where ${\sigma ^2}$ denotes the additive white Gaussian noise (AWGN) power, and $\alpha_w^2 + \alpha_{v}^2 =1$.

Then we turn our attention on calculating the outage probability of the far Rx, which is given in the following Theorem.

\begin{theorem}\label{Theorem1:Outage far UAV stochastic}
\emph{Assuming that $ \alpha _{{v}}^2 - \alpha _{w}^2{\varepsilon _{v}} > 0$, the closed-form expression in terms of outage probability of the far Rx can be expressed as}
\begin{equation}\label{outage analytical results far in theorem1}
\begin{aligned}
{\rm{P}}_v &=  1 - \frac{{3{{(m{M_v}{\sigma ^2})}^{ - \frac{3}{\alpha }}}}}{{\alpha ({D^3} - {R^3})}}\sum\limits_{n = 0}^{m - 1} {\frac{1}{{n!}}\left( {{\rm{\gamma }}\left( {n + \frac{3}{\alpha }{\rm{ + 1}},m{M_v}{\sigma ^2}{D^\alpha }} \right)  } \right.}\\
& - {\left.{ {\rm{\gamma }}\left( {n + \frac{3}{\alpha }{\rm{ + 1}},m{M_v}{\sigma ^2}{R^\alpha }} \right)} \right)} ,
\end{aligned}
\end{equation}
\emph{where ${M_v} = \frac{{{\varepsilon _v}}}{{{P_u}(\alpha _v^2 - {\varepsilon _v}\alpha _w^2)}}$, ${\varepsilon _v}=2^{R_v}-1 $, and $\gamma \left(  \cdot  \right)$ represents the lower incomplete Gamma function. }
\begin{proof}
Please refer to Appendix A.
\end{proof}
\end{theorem}

It is challenging to solve the integral in~\eqref{outage analytical results far in theorem1} directly due to the lower incomplete Gamma function. Thus, in order to gain further insights in the high SNR regime, the asymptotic behavior is analyzed, usually when the transmit SNR of the channels between the Tx-UAV and Rxs is sufficiently high, i.e., when the transmit SNR obeys ${\frac{P_u}{\sigma^2}  \to \infty }$.
\begin{corollary}\label{corollary1:Outage far UAV stochastic}
\emph{ Assuming that $\alpha _{{v}}^2 - \alpha _{w}^2{\varepsilon _{v}} > 0$, and $\frac{P_u}{\sigma^2} \to \infty$, the asymptotic outage probability of the far Rx is given by}
\begin{equation}\label{asympto v in corollary1}
\begin{aligned}
{\rm{{\hat P}}}_v &= 1 - \frac{3}{{({D^3} - {R^3})}}\sum\limits_{n = 0}^{m - 1} {\frac{{{{(m{M_v}{\sigma ^2})}^n}}}{{n!}}\frac{{{D^{\alpha n + 3}} - {R^{\alpha n + 3}}}}{{\alpha n + 3}}} \\
& + \frac{3}{{({D^3} - {R^3})}}\sum\limits_{n = 0}^{m - 1} {\frac{{{{(m{M_v}{\sigma ^2})}^{n + 1}}}}{{n!}}\frac{{{D^{\alpha (n + 1) + 3}} - {R^{\alpha (n + 1) + 3}}}}{{\alpha (n + 1) + 3}}}.
\end{aligned}
\end{equation}
\begin{proof}
Please refer to Appendix B.
\end{proof}
\end{corollary}

\begin{remark}\label{remark1:asymptotic v}
The derived results in \eqref{asympto v in corollary1} demonstrate that the outage probability of the far Rx can be decreased in the case of higher fading parameter $m$ or decreasing the target rate of the far Rx itself.
\end{remark}

\begin{remark}\label{user-centric power allocation}
Inappropriate power allocation such as, $\alpha _v^2 -  {{\varepsilon _t}}\alpha _w^2<0$, will lead to the outage probability always being one.
\end{remark}

\begin{proposition}\label{proposition1: v diversity order stochastic geo}
\emph{From \textbf{Corollary \ref{corollary1:Outage far UAV stochastic}}, one can yield the diversity order by using the high SNR approximation, and the diversity order of the far Rx in the proposed NOMA enhanced U2X frameworks is given by}
\begin{equation}\label{diversity order}
{d_{v}} =  - \mathop {\lim }\limits_{\frac{P_u}{\sigma^2}  \to \infty } \frac{{\log {\rm{\hat P}}_{v}^\infty }}{{\log \frac{P_u}{\sigma^2} }} \approx m.
\end{equation}
\end{proposition}

We then attempt to derive the outage probability for a no-fading environment by applying the limits $m \to \infty$. In this case, it is readily to derive that the small-scale fading coefficients of paired NOMA Rxs equal to one, i.e., ${{\left| {{h_{w}}} \right|}^2}={{\left| {{h_{v}}} \right|}^2}=1$. For our approach, we observe the asymptotic result of the far Rx in the following corollary.

\begin{corollary}\label{corollary: Outage far UAV stochastic m infinity}
\emph{ Assuming that $\alpha _{{v}}^2 - \alpha _{w}^2{\varepsilon _{v}} > 0$, and $m \to \infty$, the outage probability of the far Rx is given by}
\begin{equation}\label{minf final far result in corollary overall}
{\rm{P}}_v^\infty  = \left\{ \begin{array}{l}
1,\,\,{z_f} < R\\
\frac{{z_f^3 - {R^3}}}{{({D^3} - {R^3})}},\,\,R < {z_f} < D\\
0,\,\,D < {z_f}
\end{array} \right. ,
\end{equation}
where $z_f = {\left( {\frac{{{P_u}(\alpha _v^2 - {\varepsilon _v}\alpha _w^2)}}{{{\varepsilon _v}{\sigma ^2}}}} \right)^{\frac{1}{\alpha }}}$.
\begin{proof}
Please refer to Appendix C.
\end{proof}
\end{corollary}

We then turn our attention on the near Rx, and recall that the near Rx needs to decode the signal for the far Rx before decoding its own message via SIC, and the SINR can be given by
\begin{equation}\label{SINR stochastic near w}
\begin{aligned}
{\rm{P}}_w &= {\rm{P}}\left( {\log \left( {1 + \frac{{{P_u}{{\left| {{h_w}} \right|}^2}d_w^{ - \alpha }\alpha _v^2}}{{{\sigma ^2} + {P_u}{{\left| {{h_w}} \right|}^2}d_w^{ - \alpha }\alpha _w^2}}} \right) < {R_v}} \right)  \\
& +{\rm{P}}\left( {\log \left( {1 + \frac{{{P_u}{{\left| {{h_w}} \right|}^2}d_w^{ - \alpha }\alpha _v^2}}{{{\sigma ^2} + {P_u}{{\left| {{h_w}} \right|}^2}d_w^{ - \alpha }\alpha _w^2}}} \right) > {R_v}, } \right. \\
&\left. {\log \left( {1 + \frac{{{P_u}{{\left| {{h_w}} \right|}^2}d_w^{ - \alpha }\alpha _w^2}}{{{\sigma ^2}}}} \right) < {R_w}} \right).
\end{aligned}
\end{equation}
Then, the outage probability of the near Rx can be derived in the following Theorem.

\begin{theorem}\label{Theorem2:Outage near UAV stochastic}
\emph{Assuming that $ \alpha _{{v}}^2 - \alpha _{w}^2{\varepsilon _{v}} > 0$, the closed-form expression in terms of outage probability of the near Rx can be expressed as}
\begin{equation}\label{outage analytical results near in theorem2_NOMA}
\begin{aligned}
{\rm{P}}_w &=  1 - \frac{{3{{(m{M_{w*}}{\sigma ^2})}^{ - \frac{3}{\alpha }}}}}{{\alpha ({R^3} - {r_0^3})}}\sum\limits_{n = 0}^{m - 1} {\frac{1}{{n!}}\left( {{\rm{\gamma }}\left( {n + \frac{3}{\alpha }{\rm{ + 1}},m{M_{w*}}{\sigma ^2}{R^\alpha }} \right) }\right.}\\
& {\left.{- {\rm{\gamma }}\left( {n + \frac{3}{\alpha }{\rm{ + 1}},m{M_{w*}}{\sigma ^2}{r_0^\alpha }} \right)} \right)} .
\end{aligned}
\end{equation}
\emph{where ${M_w} = \frac{{{\varepsilon _w}}}{{{P_u}\alpha _w^2}}$, $M_{w*}=max\{M_v, M_w \}$, and $\varepsilon _w=2^{R_w}-1$. }
\begin{proof}
Based on the SINR analysis in \eqref{SINR stochastic near w}, and following the similar procedure in Appendix A, with interchanging ${M_{v}}$ with ${M_{w*}}$, we can obtain the desired result in \eqref{outage analytical results near in theorem2_NOMA}. Thus, the proof is complete.
\end{proof}
\end{theorem}

Based on the results in~\eqref{outage analytical results near in theorem2_NOMA}, we can derive the asymptotic result of the near Rx in the following corollary.

\begin{corollary}\label{corollary4: Outage near UAV stochastic}
\emph{ Assuming that $\alpha _{{v}}^2 - \alpha _{w}^2{\varepsilon _{v}} > 0$, and $\frac{P_u}{\sigma^2} \to \infty$, the asymptotic outage probability of the near Rx is given by}
\begin{equation}\label{asympto w in corollary2}
\begin{aligned}
{\rm{\hat P}}_w & = 1 - \frac{3}{{({R^3} - {r_0^3})}}\sum\limits_{n = 0}^{m - 1} {\frac{{{{(m{M_v}{\sigma ^2})}^n}}}{{n!}}\frac{{{R^{\alpha n + 3}} - r_0^{\alpha n + 3}}}{{\alpha n + 3}}} \\
& + \frac{3}{{({R^3} - {r_0^3})}}\sum\limits_{n = 0}^{m - 1} {\frac{{{{(m{M_v}{\sigma ^2})}^{n + 1}}}}{{n!}}\frac{{{R^{\alpha (n + 1) + 3}} - r_0^{\alpha (n + 1) + 3}}}{{\alpha (n + 1) + 3}}}.
\end{aligned}
\end{equation}
\begin{proof}
Please refer to Appendix B, the asymptotic outage probability of the near Rx can be readily proved.
\end{proof}
\end{corollary}

\begin{remark}\label{remark2:diversity orders stochastic}
Following steps similar to the proof in \textbf{Proposition 1}, the diversity order of the near Rx can be obtained, which is also $m$.
\end{remark}

It is also worth estimating the outage probability of the near Rx in the no-fading environment by applying the limits $m \to \infty$. Thus, for our approach, we observe the asymptotic result of the near Rx in the following corollary.

\begin{corollary}\label{corollary: Outage near UAV stochastic m infinity}
\emph{ Assuming that $\alpha _{{v}}^2 - \alpha _{w}^2{\varepsilon _{v}} > 0$, and $m \to \infty$, the outage probability of the near Rx is given by}
\begin{equation}\label{minf final near result in corollary overall}
{\rm{P}}_w^\infty  = \left\{ \begin{array}{l}
1,\,\,{z_{n*}} < r_0\\
\frac{{z_{n*}^3 - r_0^3}}{{({R^3} - r_0^3)}},\,\,r_0 < {z_{n*}} < R\\
0,\,\,R < {z_{n*}}
\end{array} \right.,
\end{equation}
where ${z_n} = {\left( {\frac{{{P_u}\alpha _w^2}}{{{\varepsilon _w}{\sigma ^2}}}} \right)^{\frac{1}{\alpha }}}$, ${z_{n*}} = {\rm{Min}}\left\{ {{z_n},{z_f}} \right\}$.
\begin{proof}
Similar to Appendix C, with interchanging $z_f$ with $z_{n*}$, we can obtain the desired result in~\eqref{minf final near result in corollary overall}, and the proof can is complete.
\end{proof}
\end{corollary}

In order to provide more insights for U2X frameworks, the outage probability of the Rxs is also derived in the OMA case, i.e., TDMA. We propose two possible scenarios for the OMA case, where a user is uniformly located in the sphere $\mathcal{V}^3$ for the first scenario. Thus, on the one hand, the outage probability of the OMA case can be given by
\begin{equation}\label{Outage OMA case onlyone}
{\rm P}_{o} =  {\mathbb{E}} \left( {\log \left( {1 + \frac{{P_u {{\left| {{h_{o}}} \right|}^2}d_o^{-\alpha}}}{{\sigma^2   }}} \right) < {R_{o}}} \right),
\end{equation}
where ${h_{o}}$ denotes the small-scale fading in the first OMA scenario, and thus the outage probability can be derived in the following Theorem.

\begin{theorem}\label{Theorem3:Outage onlyone OMA UAV stochastic}
\emph{The outage probability of the Rx in the first OMA scenario can be expressed as}
\begin{equation}\label{outage exact expression onlyone OMA in theorem3}
\begin{aligned}
{\rm{P}}_O^1& = 1 - \frac{{3{{(m{M_o}{\sigma ^2})}^{ - \frac{3}{\alpha }}}}}{{\alpha ({D^3} - r_0^3)}}\sum\limits_{n = 0}^{m - 1} {\frac{1}{{n!}}\left( {\gamma \left( {n + \frac{3}{\alpha }+1,m{M_o}{\sigma ^2}{D^\alpha }} \right) }\right.} \\
&{\left.{ - \gamma \left( {n + \frac{3}{\alpha }+1,m{M_o}{\sigma ^2}r_0^\alpha } \right)} \right)},
\end{aligned}
\end{equation}
\emph{where ${M_o} = \frac{{{\varepsilon _o}}}{{{P_u}}}$, $R_o$ denotes the target rate of the OMA Rx, and $\varepsilon_o=2^{2R_o}-1$. }
\begin{proof}
Based on the SINR analysis in \eqref{Outage OMA case onlyone}, and following the similar procedures in Appendix A, with interchanging ${M_{v}}$ with ${M_{o}}$, we can obtain the desired result in~ \eqref{outage exact expression onlyone OMA in theorem3}. Thus, the proof is complete.
\end{proof}
\end{theorem}

On the other hand, another OMA scenario is also worth estimating, where two OMA Rxs, near Rx and far Rx are located in the small sphere $\mathcal{V}_1^3$ and large hollow sphere $\mathcal{V}_2^3$ with radius $R$ and $D$ ($D > R$), respectively. The outage probability for the second OMA scenario can be derived in the following Theorem.

\begin{theorem}\label{Theorem4:Outage two user OMA near}
\emph{The outage probability of both near and far Rxs in the second OMA scenario can be expressed as}
\begin{equation}\label{outage exact expression of near user in two user OMA in theore4}
\begin{aligned}
&{\rm{P}}_{O,w}^2 =  1 - \frac{{3{{(m{M_{o,w}}{\sigma ^2})}^{ - \frac{3}{\alpha }}}}}{{2\alpha ({R^3} - r_0^3)}}\sum\limits_{n = 0}^{m - 1} {\frac{1}{{n!}}}\\
& \times { \left( {\gamma \left( {n + \frac{3}{\alpha }+1,m{M_{o,w}}{\sigma ^2}{R^\alpha }} \right) - \gamma \left( {n + \frac{3}{\alpha }+1,m{M_{o,w}}{\sigma ^2}r_0^\alpha } \right)} \right)},
\end{aligned}
\end{equation}
and
\begin{equation}\label{outage exact expression of far user in two user OMA in theore4}
\begin{aligned}
&{\rm{P}}_{O,v}^2 = 1 - \frac{{3{{(m{M_{o,v}}{\sigma ^2})}^{ - \frac{3}{\alpha }}}}}{{2\alpha ({D^3} - R^3)}}\sum\limits_{n = 0}^{m - 1} {\frac{1}{{n!}}}\\
&\times {\left( {\gamma \left( {n + \frac{3}{\alpha }+1,m{M_{o,v}}{\sigma ^2}{D^\alpha }} \right) - \gamma \left( {n + \frac{3}{\alpha }+1,m{M_{o,v}}{\sigma ^2}R^\alpha } \right)} \right)},
\end{aligned}
\end{equation}
where ${M_{o,v}} = \frac{{{\varepsilon _{o,v}}}}{{{P_u}}}$, ${M_{o,w}} = \frac{{{\varepsilon _{o,w}}}}{{{P_u}}}$, $R_{o,w}$ and $R_{o,v}$ denote the target rates of the near and far OMA Rx, $\varepsilon_{o,v}=2^{2R_{o,v}}-1$, and $\varepsilon_{o,w}=2^{2R_{o,w}}-1$.

\begin{proof}
Following the similar procedure in Theorem~\ref{Theorem3:Outage onlyone OMA UAV stochastic}, we can obtain the desired result in~\eqref{outage exact expression of near user in two user OMA in theore4} and~\eqref{outage exact expression of far user in two user OMA in theore4}. Thus, the proof is complete.
\end{proof}
\end{theorem}

Since Rx can receive three groups of signals including LoS, strong reflected NLoS signals, and multiple reflected components which cause multi-path fading. One common approach for modeling A2G propagation channel is to consider LoS and NLoS components along with their occurrence probabilities separately as shown in~\cite{PLoS_model}. Therefore, in order to provide more engineering insights, and based on the model in~\cite{Saad_D2D_UAV,PLoS_model}, a probability of LoS scenario is also provided in the following Proposition.

\begin{proposition}\label{corollary: PLos model Outage near UAV stochastic}
\emph{ Depending on the occurrence probabilities of LoS propagations, the outage probability conditioned on the probability of LoS of paired NOMA Rxs can be given by}
\begin{equation}\label{minf final near result in corollary overall}
P_w^o = {P_{{\rm{LoS}}}}P_{w{\rm{,LoS}}} + (1 - {P_{{\rm{LoS}}}})P_{w{\rm{,NLoS}}},
\end{equation}
and
\begin{equation}\label{minf final far result in corollary overall}
P_v^o = {P_{{\rm{LoS}}}}P_{v{\rm{,LoS}}} + (1 - {P_{{\rm{LoS}}}})P_{v{\rm{,NLoS}}},
\end{equation}
where ${P_{{\rm{LoS}}}}$ denotes the LoS probability, $P_{w{\rm{,LoS}}}$ and $P_{w{\rm{,NLoS}}}$ denote the outage probability of the LoS scenario and NLoS scenario, respectively.
\begin{proof}
Depending on the LoS and NLoS connection between Tx-UAV and Rx, the received signal power at the near Rx is given by
\begin{equation}\label{PloS seperate}
{P_w} = \left\{ \begin{array}{l}
{P_u}{L_w}{\left| {{h_{w,m = 1}}} \right|^2},{\rm{NLoS}}\\
{P_u}{L_w}{\left| {{h_{w,m > 1}}} \right|^2},{\rm{LoS}}
\end{array} \right.,
\end{equation}
where ${\left| {{h_{w,m = 1}}} \right|}$ and ${\left| {{h_{w,m > 1}}} \right|}$  follow the distribution in~\eqref{channel matrix,eq3} with fading parameter $m=1$ and $m>1$, respectively. Thus, the result in~\eqref{minf final near result in corollary overall} and \eqref{minf final far result in corollary overall} can be readily proved.
\end{proof}
\end{proposition}

\subsection{Ergodic Rate}
In the U2X frameworks, the ergodic rate is a critical metric, which is worth estimating for performance evaluation. Therefore, we focus on analyzing the ergodic rates of individual U2X Rxs, which are determined by their channel conditions and geometry parameters in the proposed framework. The asymptotic ergodic rate for the near Rx is shown in the following corollary.

\begin{corollary}\label{corollary3:ergodic rate near UAV stochastic}
\emph{ The achievable ergodic rate of the near Rx can be expressed as follows:}
\begin{equation}\label{asympto w erogodic rate in corollary3}
\begin{aligned}
{R_w} =& \frac{3}{{\alpha \ln \left( 2 \right)\left( {{R^3} - r_0^3} \right)}}\sum\limits_{n = 0}^{m - 1} {\frac{1}{{n!}}} \sum\limits_{k = 0}^\infty  {\frac{{\Gamma \left( {{\phi _2} - 1} \right)\Gamma \left( {n + 1 + k} \right)}}{{\Gamma \left( {{\phi _2} + k} \right)}} }\\
&\times {C^{n + k}}\left( {{R^{{\phi _1}}}\exp (C{R^\alpha })\Gamma \left( { - k - n,C{R^\alpha }} \right) } \right.\\
& \left. { - r_0^{{\phi _1}}\exp (Cr_0^\alpha )\Gamma \left( { - k - n,Cr_0^\alpha } \right)} \right),
\end{aligned}
\end{equation}
where ${\phi _1} = \alpha n + 3 + \alpha k$, ${\phi _2} = n + \frac{3}{\alpha } + 1$, $C = \frac{{m{\sigma ^2}}}{{{P_u}\alpha _w^2}}$, and $\Gamma \left( , \right)$ denotes upper incomplete Gamma function.
\begin{proof}
Please refer to Appendix D.
\end{proof}
\end{corollary}

The ergodic rate of the far Rx is also worth evaluating in the following corollary.

\begin{corollary}\label{corollary3:ergodic rate far UAV stochastic}
\emph{ The achievable ergodic rate of the far Rx can be expressed as follows:}
\begin{equation}\label{asympto v erogodic rate in corollary4}
\begin{aligned}
{R_v} =& {\log _2}(1 + \frac{{\alpha _v^2}}{{\alpha _w^2}}) + {Q_1}{\log _2}(1 + \frac{{\alpha _v^2}}{{\alpha _w^2}}) - {Q_2}{\log _2}(1 + \frac{{\alpha _v^2}}{{\alpha _w^2}}),
\end{aligned}
\end{equation}
where ${Q_1} = \frac{{3({D^{\alpha n + 3}} - {R^{\alpha n + 3}})}}{{({D^3} - {R^3})(\alpha n + 3)}}\sum\limits_{n = 0}^{m - 1} {\frac{1}{{n!}}} {\left( {\frac{{m{\sigma ^2}}}{{{P_u}}}} \right)^n}$, and ${Q_2} = \frac{{3({D^{\alpha (n + 1) + 3}} - {R^{\alpha (n + 1) + 3}})}}{{({D^3} - {R^3})(\alpha (n + 1) + 3)}}\sum\limits_{n = 0}^{m - 1} {\frac{1}{{n!}}} {\left( {\frac{{m{\sigma ^2}}}{{{P_u}}}} \right)^{(n + 1)}}$.
\begin{proof}
Similar to Appendix~\ref{Appendix:Ds}, by interchanging the upper bound of the intergral, the result in~\eqref{asympto v erogodic rate in corollary4} can be readily proved.
\end{proof}
\end{corollary}

\begin{remark}\label{remark4:ergodic rate far}
It is proved that the ergodic rates of the far Rxs is entirely dependent on the power allocation factors in the high SNR regime, which is equal to ${\rm{log}}_2(1+\frac{\alpha_v^2}{\alpha_w^2})$.
\end{remark}

\begin{remark}\label{remark3:ergodic rate near}
By utilizing the exponential series expansion to the upper incomplete Gamma function, the high SNR slope of the proposed framework for any fading parameters can be readily derived, which are 1 and 0 for near Rxs and far Rxs, respectively.
\end{remark}

In order to provide the benchmark of U2X frameworks, we also derive the asymptotic results of achievable ergodic rate in the case of OMA enhanced U2X frameworks in the following corollary.

\begin{corollary}\label{corollary4:ergodic rate near UAV stochastic OMA case}
\emph{ The achievable ergodic rate of the Rx in the first OMA scenario can be expressed as follows:}
\begin{equation}\label{asympto w erogodic rate in corollary4 OMA case}
\begin{aligned}
{R_o^1} =& \frac{3}{{\alpha \ln \left( 2 \right)\left( {{D^3} - r_0^3} \right)}}\sum\limits_{n = 0}^{m - 1} {\frac{1}{{n!}}} \sum\limits_{k = 0}^\infty  {\frac{{\Gamma \left( {{\phi _2} - 1} \right)\Gamma \left( {n + 1 + k} \right)}}{{\Gamma \left( {{\phi _2} + k} \right)}} }\\
&\times {C_o}^{n + k}\left( {{D^{{\phi _1}}}\exp ({C_o}{D^\alpha })\Gamma \left( { - k - n,{C_o}{D^\alpha }} \right) } \right.\\
&\left.{ - r_0^{{\phi _1}}\exp ({C_o}r_0^\alpha )\Gamma \left( { - k - n,{C_o}r_0^\alpha } \right)} \right) ,
\end{aligned}
\end{equation}
where ${C_o} = \frac{{m{\sigma ^2}}}{{{P_u}}}$.
\begin{proof}
In the case of OMA enhanced U2X frameworks, the achievable ergodic rate can be defined as
\begin{equation}\label{defination erogodic rate in corollary4 proof OMA case}
\begin{aligned}
{R_o^1} = \mathbb{E} \left\{ {{{\log }_2}\left( {1 + SIN{R_o}\left( {{x_o}} \right)} \right)} \right\}.
\end{aligned}
\end{equation}
Similar to the steps from \eqref{w ergodic exp} to \eqref{expression ergodic rate}, the ergodic rate can be obtained as
\begin{equation}\label{expression erogodic rate in corollary4 proof OMA case}
\begin{aligned}
&{R_o^1} = \frac{3}{({{D^3} - r_0^3}){\ln \left( 2 \right)}} \\
& \times \int\limits_0^\infty  {\frac{{\sum\limits_{n = 0}^{m - 1} {\frac{{{{\left( {{C_o}x} \right)}^n}}}{{n!}}} \int_{{r_0}}^D {{r^{\alpha n + 2}}\exp \left( { - {C_o}x{r^\alpha }} \right)} dr}}{{1 + x}}} dx.
\end{aligned}
\end{equation}
Again, similar to the steps from \eqref{full express} to \eqref{J2_appendixC}, the results in \eqref{asympto w erogodic rate in corollary4 OMA case} can be obtained.
\end{proof}
\end{corollary}

Then, the ergodic rate of Rxs in the second OMA scenario can be derived in the following Corollary.
\begin{corollary}\label{corollary5:ergodic rate near second OMA case}
\emph{ The achievable ergodic rate of the Rxs in the second OMA scenario can be expressed as follows:}
\begin{equation}\label{erogodic rate near in corollary5 second OMA case}
\begin{aligned}
{R_{o,w}^2} =& \frac{3}{{2\alpha \ln \left( 2 \right)\left( {{R^3} - r_0^3} \right)}}\sum\limits_{n = 0}^{m - 1} {\frac{1}{{n!}}} \sum\limits_{k = 0}^\infty  {\frac{{\Gamma \left( {{\phi _2} - 1} \right)\Gamma \left( {n + 1 + k} \right)}}{{\Gamma \left( {{\phi _2} + k} \right)}} }\\
&\times {C_o}^{n + k}\left( {{R^{{\phi _1}}}\exp ({C_o}{R^\alpha })\Gamma \left( { - k - n,{C_o}{R^\alpha }} \right) }\right. \\
& \left.{- r_0^{{\phi _1}}\exp ({C_o}r_0^\alpha )\Gamma \left( { - k - n,{C_o}r_0^\alpha } \right)} \right) ,
\end{aligned}
\end{equation}
and
\begin{equation}\label{erogodic rate far in corollary5 second OMA case}
\begin{aligned}
{R_{o,v}^2} =& \frac{3}{{2\alpha \ln \left( 2 \right)\left( {{D^3} - R^3} \right)}}\sum\limits_{n = 0}^{m - 1} {\frac{1}{{n!}}} \sum\limits_{k = 0}^\infty  {\frac{{\Gamma \left( {{\phi _2} - 1} \right)\Gamma \left( {n + 1 + k} \right)}}{{\Gamma \left( {{\phi _2} + k} \right)}} }\\
&\times {C_o}^{n + k}\left( {{D^{{\phi _1}}}\exp ({C_o}{D^\alpha })\Gamma \left( { - k - n,{C_o}{D^\alpha }} \right) } \right.\\
&- \left. { R^{{\phi _1}}\exp ({C_o}R^\alpha )\Gamma \left( { - k - n,{C_o}R^\alpha } \right)} \right).
\end{aligned}
\end{equation}
\begin{proof}
Similar to the steps from \eqref{full express} to \eqref{J2_appendixC}, the results in \eqref{erogodic rate near in corollary5 second OMA case} and \eqref{erogodic rate far in corollary5 second OMA case} can be obtained.
\end{proof}
\end{corollary}

\subsection{Spectrum Efficiency}

Based on the analytical results of last two subsections, the spectrum efficiency of the proposed framework can be given in the following Proposition.

\begin{proposition}\label{spectrum efficiency NOMA proposition}
\emph{In the high SNR regime, the spectrum efficiency of the proposed NOMA enhanced U2X frameworks is}
\begin{equation}\label{spectrum efficiency of NOMA}
{\tau _N} = {\log _2}(1 + \frac{{\alpha _v^2}}{{\alpha _w^2}}) + {R_w},
\end{equation}
where ${R_w}$ is obtained from \eqref{asympto w erogodic rate in corollary3}.
\end{proposition}

We also want to derive the performance gap between NOMA and OMA enhanced U2X frameworks from the perspective of spectrum efficiency, which is given in the following Proposition.
\begin{proposition}\label{spectrum efficiency gap}
\emph{In the high SNR regime, the spectrum efficiency gap of the proposed NOMA enhanced U2X frameworks is}
\begin{equation}\label{spectrum efficiency gap}
{\tau} ={\tau _N}- {R_o^1},
\end{equation}
or
\begin{equation}\label{spectrum efficiency gap}
{\tau} ={\tau _N}- {R_{o,w}^2}-{R_{o,v}^2},
\end{equation}
where ${R_o^1}$ is obtained from~\eqref{asympto w erogodic rate in corollary4 OMA case}, ${R_{o,w}^2}$ is obtained from~\eqref{erogodic rate near in corollary5 second OMA case}, ${R_{o,v}^2}$ is obtained from~\eqref{erogodic rate far in corollary5 second OMA case}.
\end{proposition}

\section{Numerical Studies}

In this section, numerical results are provided to facilitate the performance evaluation of NOMA enhanced U2X frameworks. Monte Carlo simulations are conducted to verify analytical results. In the considered network, it is assumed that the power allocation factors are $\alpha_{v}^2=0.6$ for the far Rx and $\alpha_{w}^2=0.4$ for the near Rx. The power of AWGN noise is set as $\sigma^2= −-90$ dBm. It is also worth noting that LoS and NLoS scenarios are indicated by the Nakagami fading parameter $m$, where $m = 1$ for NLoS scenarios (Rayleigh fading) and $m >1$ for LoS scenarios. Without loss of generality, we use $m=2, 3$ to represents LoS scenario in Section IV. In order to avoid infinite received power, the minimum distance $r_0=1m$.

\subsection{Outage Probabilities}

\begin{figure*}[t!]
\centering
\includegraphics[width =4in]{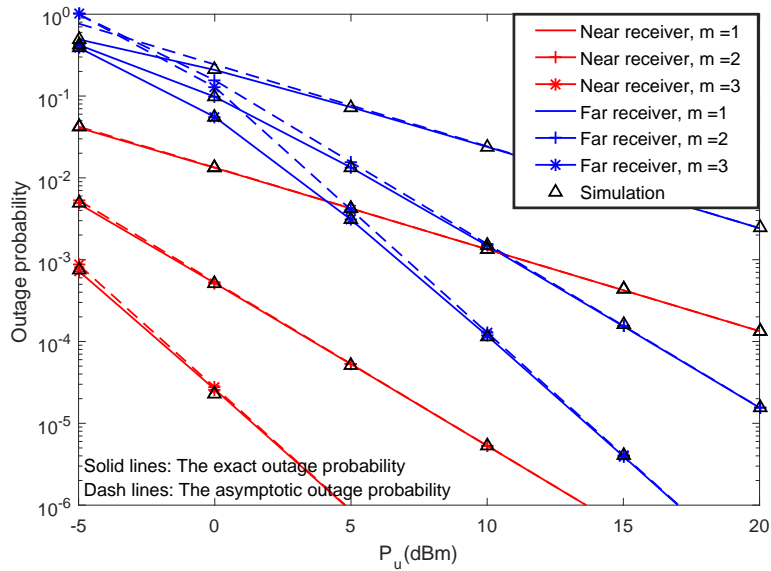}
\caption{Outage probability of NOMA enhanced U2X framework versus transmit power in both NLoS and LoS scenarios, where the fading parameters $m=1, 2, 3$. The target rate of near Rxs and far Rxs are $R_w=1.5$ bit per channel use (BPCU) and $R_v=1$ BPCU, respectively. The path loss exponent $\alpha=4$. The exact results are calculated from~\eqref{outage analytical results far in theorem1} and~\eqref{outage analytical results near in theorem2_NOMA}. The asymptotic results are derived from~\eqref{asympto v in corollary1} and~\eqref{asympto w in corollary2}.}
\label{Outage NOMA U2U fig 2}
\end{figure*}

\emph{1) Impact of Fading:} In Fig.~\ref{Outage NOMA U2U fig 2}, we evaluate the outage probability of paired NOMA Rxs in both NLoS and LoS scenarios. The solid curves, dashed curves and dotted curves are the exact results, analytical results and asymptotic results, respectively. We can see that, as the power of the UAV increases, the outage probability of both near and far NOMA Rxs decreases.
This is due to the fact that, as higher transmit power level of the UAV is deployed, the received SINR improves. It is also confirmed the close agreement between the simulation and analytical results in the high SNR regime, which verifies our analytical results. Note that the slope of curves for both paired NOMA Rxs is $m$, which verifies that the diversity orders of schemes are $m$. This phenomenon validates the insights from \textbf{Proposition~\ref{proposition1: v diversity order stochastic geo}} and \textbf{Remark~\ref{remark2:diversity orders stochastic}}.

\begin{figure*}[t!]
\centering
\includegraphics[width =4in]{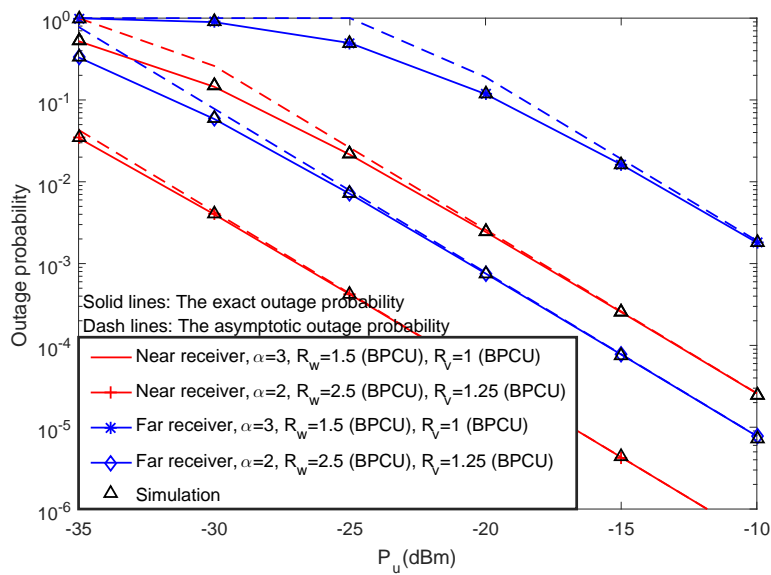}
\caption{Outage probability of NOMA Rxs with different path loss exponent, where the fading parameters $m=2$.}
\label{Outage NOMA diff_alpha}
\end{figure*}

\emph{2) Impact of Path-Loss Exponent:} We study the impact of path-loss exponent $\alpha$ and target rate $R_w$ and $R_v$ on outage probability in Fig.~\ref{Outage NOMA diff_alpha}. It can be observed that the outage probability decreases in the case of lower path-loss exponent. Note that the free space model, where the path loss exponent $\alpha=2$, is also evaluated to provide more engineering insights.
It is also worth noting that in Fig.~\ref{Outage NOMA diff_alpha}, the accuracy of {\textbf{Corollary~\ref{corollary1:Outage far UAV stochastic}} and {\textbf{Corollary~\ref{corollary4: Outage near UAV stochastic}} can be confirmed similarly.

\emph{3) Outage Sum Rate:} Fig.~\ref{Outage sum rate} plots the system outage sum rate versus transmit power with different targeted rates and fading parameters. One can observe that the case $m=\infty$ achieves the highest throughput since it has the lowest outage probability among the three selection fading parameters. The figure also demonstrates the existence of the throughput ceilings in the high SNR region. This is due to the fact that the outage probability is approaching zero and the throughput is determined only by the targeted data rate. It is also worth mentioning that for the case of $\alpha=2$ and $m \to \infty$, the channel model of the proposed U2X framework can be recognized as free space model.

\begin{figure*}[t!]
\centering
\includegraphics[width =4in]{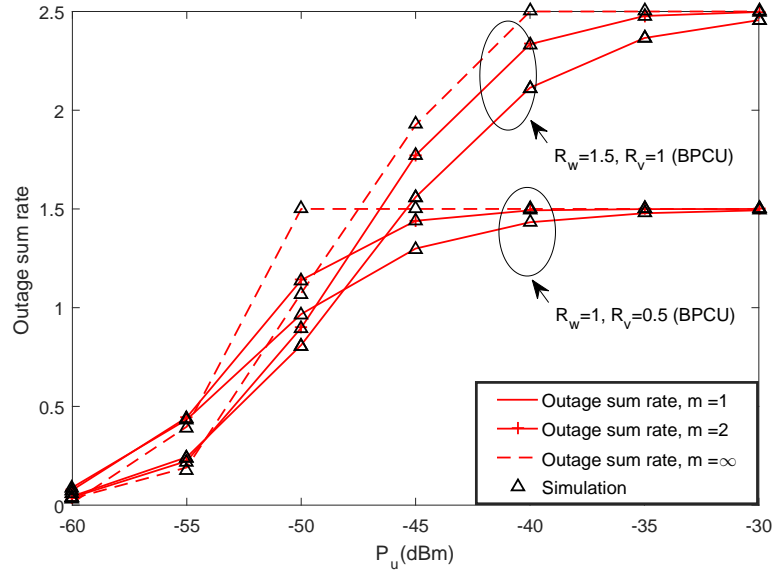}
\caption{Outage sum rate of NOMA Rxs with different rates, where the fading parameters $m=1, 2, \infty$. The path loss exponent $\alpha=2$. The outage sum rates of NOMA enhanced U2X framework is derived by $(1-{\rm{P}}_v)R_v+(1-{\rm{P}}_w)R_w$.}
\label{Outage sum rate}
\end{figure*}

\emph{3) Probability of LoS:} Fig.~\ref{PLoS model} plots the outage probability versus transmit power with different LoS probabilities. The outage probability for the cases of $m=1$ and $m=3$ are plotted as the benchmark schemes. One can readily observe that the outage probability of paired NOMA Rxs mainly depends on the NLoS case even in the case of ${\rm{P_{LoS}}}=0.8$. We can also see that the diversity order of the proposed probability of the probability of LoS model is one, which indicates that for the case of fixed LoS probability, the outage performance of paired NOMA Rxs mainly depends on the user with poor channel condition.

\begin{figure*}[t!]
\centering
\includegraphics[width =4in]{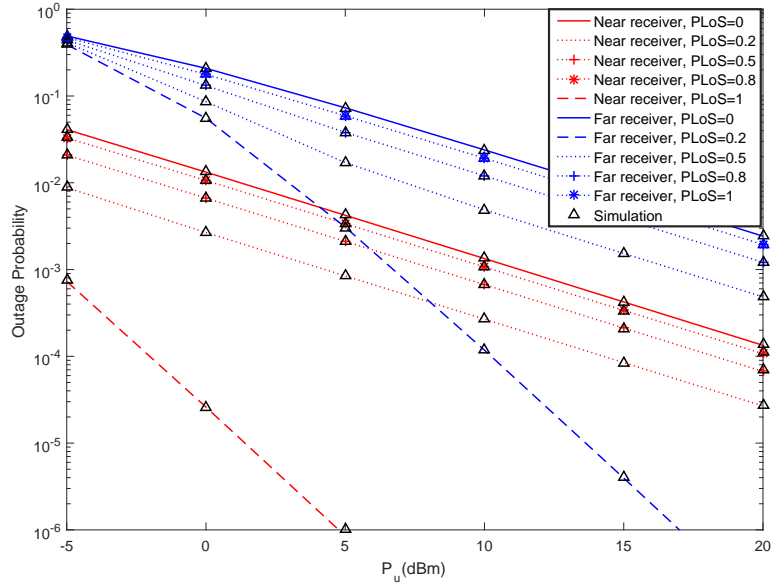}
\caption{Outage probability of paired NOMA Rxs with different LoS probability, where the NLoS case is denoted by $m=1$, and LoS case is denoted by $m=3$. The results are derived from~\eqref{minf final near result in corollary overall} and~\eqref{minf final far result in corollary overall}.}
\label{PLoS model}
\end{figure*}

\begin{figure*}[t!]
\centering
\includegraphics[width =4in]{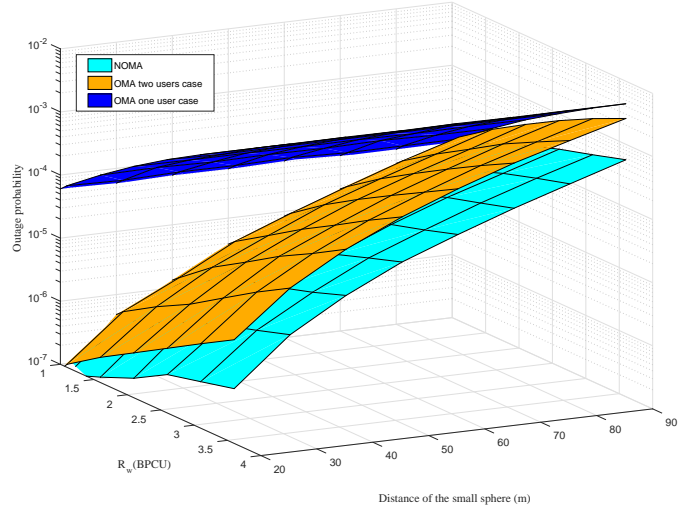}
\caption{Outage probability of paired NOMA Rxs and OMA Rxs, where the fading parameters $m=2$. The target rate of the far user is $R_v=0.5$ BPCU. The path loss exponent $\alpha=4$.}
\label{Outage NOMA 3D}
\end{figure*}

\emph{4) Performance with OMA:} In Fig.~\ref{Outage NOMA 3D}, we evaluate the system outage probability in both NOMA and OMA scenarios versus the target rate of near Rxs and the distance of the small sphere. The outage probability of NOMA enhanced U2X frameworks is derived by ${\rm{{\hat P}}}_v \times {\rm{{\hat P}}}_w $. The two users scenario and one user scenario of OMA enhanced U2X frameworks in terms of outage probability are derived by ${\rm{P}}_{O,w}^2 \times {\rm{P}}_{O,v}^2$ and ${{\rm{P}}_{O}^1} \times {{\rm{P}}_{O}^1}$, respectively. As can be seen from Fig.~\ref{Outage NOMA 3D}, the outage probability of NOMA enhanced U2X framework is lower than the OMA enhanced U2X frameworks, which implies that NOMA enhanced U2X frameworks is capable of providing better access services than OMA.

\subsection{Ergodic Rates}

\emph{5) Impact of fading:} Fig.~\ref{Ergodic rate NOMA U2U fig 3} compares the ergodic rates of paired NOMA Rxs versus transmit power with different fading parameters. Several observations can be drawn as follows: 1) An ergodic rate ceiling for far Rxs exists even if the transmit power goes to infinity. This is because that the power allocation factors are the dominant components of far Rxs in terms of ergodic rate. 2) The solid curves, dashed curves and triangles show the precise agreement between the exact results, asymptotic results and simulations, which verify our results. 3) As can be seen from the figure, the high SNR slope of the near Rxs is one, which also verifies \textbf{Remark~\ref{remark3:ergodic rate near}}. 4) The ergodic rate of near Rxs with LoS link is higher than the NLoS case. This is because the LoS propagation increases the received power level, which increases the ergodic rate of near Rxs.

\begin{figure*}[t!]
\centering
\includegraphics[width =4in]{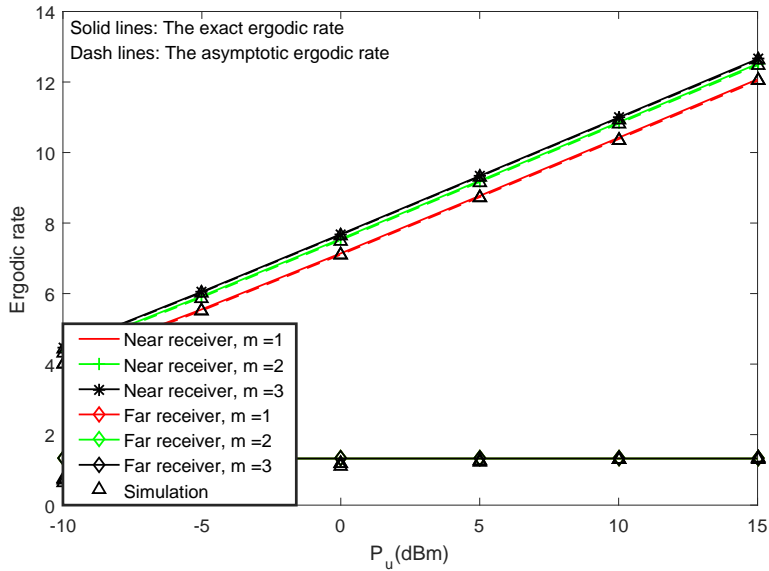}
\caption{Ergodic rate of NOMA enhanced U2X framework versus transmit power in both NLoS and LoS scenarios, where the fading parameters $m=1, 2, 3$. The path loss exponent $\alpha=4$. The asymptotic results are derived from \eqref{asympto w erogodic rate in corollary3}.}
\label{Ergodic rate NOMA U2U fig 3}
\vspace{-0.2in}
\end{figure*}

\emph{6) Spectrum Efficiency:} Fig.~\ref{spectrum_efficiency_fig 6} plots the spectrum efficiency of the proposed U2X frameworks with NOMA and OMA versus transmit power. The curves representing the performance of NOMA enhanced U2X frameworks are from \eqref{spectrum efficiency of NOMA}. The performance of OMA enhanced U2X frameworks is illustrated as a benchmark to demonstrate the
effectiveness of our proposed framework. It is can be observed that the spectrum efficiency of U2X frameworks improves as the transmit power increases. It is also worth noting that the performance of NOMA enhanced U2X frameworks outperforms the conventional OMA enhanced U2X frameworks, which in turn enhances the spectrum efficiency of the whole frameworks.

\begin{figure*}[t!]
\centering
\includegraphics[width =4in]{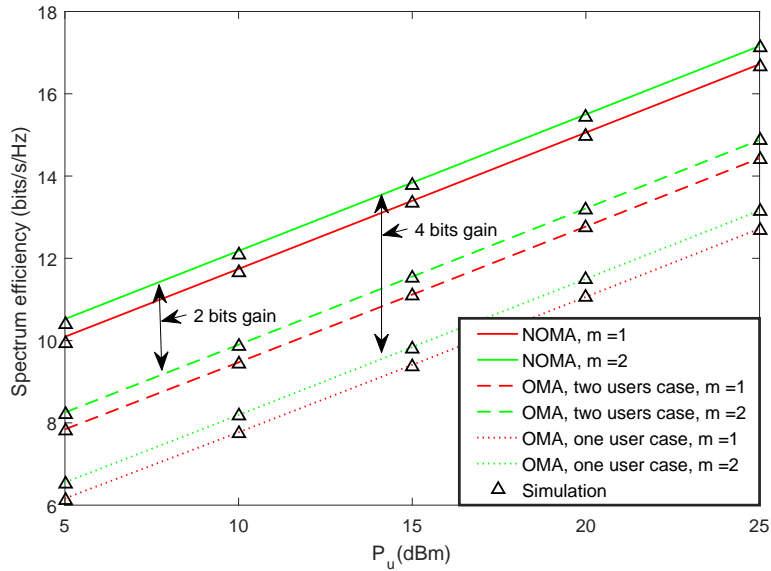}
\caption{Spectrum efficiency of both NOMA and OMA enhanced U2X frameworks versus transmit power in both NLoS and LoS scenarios, where the fading parameters $m=1, 2$. The asymptotic results are derived from \eqref{asympto w erogodic rate in corollary3}.}
\label{spectrum_efficiency_fig 6}
\end{figure*}

\emph{7) High SNR Slope:} Fig.~\ref{high SNR slope_fig 7} plots the high SNR slope of paired NOMA Rxs versus transmit power. It is observed that the high SNR slope of far Rxs and near Rxs goes to zero and one, respectively. This behavior can be explained as follows. The ergodic rate of far Rxs, which changes slightly in the high SNR regime, is entirely determined by the power allocation factors. Thus, in the low SNR regime, as transmit power $P_u$ increases, the high SNR slope of far Rxs is increased. For near Rxs, the SNR slope increases monotonously, which shows the high SNR slope for near Rxs is one in the high SNR regime. Another insight is that the LoS link accelerates the increasing rate and the decreasing rate of the paired NOMA Rxs. As shown in TABLE~\ref{DIVERSITY ORDER AND HIGH SNR SLOPE FOR U2X Networks}, the diversity orders and high SNR slopes of paired Rxs for both NOMA and OMA enhanced U2X framework are summarized to illustrate the comparison between them. In TABLE~\ref{DIVERSITY ORDER AND HIGH SNR SLOPE FOR U2X Networks}, we use “D” and “S” to represent the diversity order and high SNR slope, respectively.

\begin{table}
\caption{\\ DIVERSITY ORDER AND HIGH SNR SLOPE FOR U2X FRAMEWORKS}
\centering
\begin{tabular}{|c|c|c|c|}
\hline
Access Mode & Rx & D & S \\
\hline
\multirow{2}{*}{ NOMA}
& Near & m & 1 \\
\cline{2-4}
& Far & m & 0 \\
\hline
\multirow{3}{*}{OMA}
& Near & m & 0.5 \\
\cline{2-4}
& Far & m & 0.5 \\
\cline{2-4}
& Only one & m & 1 \\
\hline
\end{tabular}
		
\label{DIVERSITY ORDER AND HIGH SNR SLOPE FOR U2X Networks}
\end{table}

\begin{figure*}[t!]
\centering
\includegraphics[width =4in]{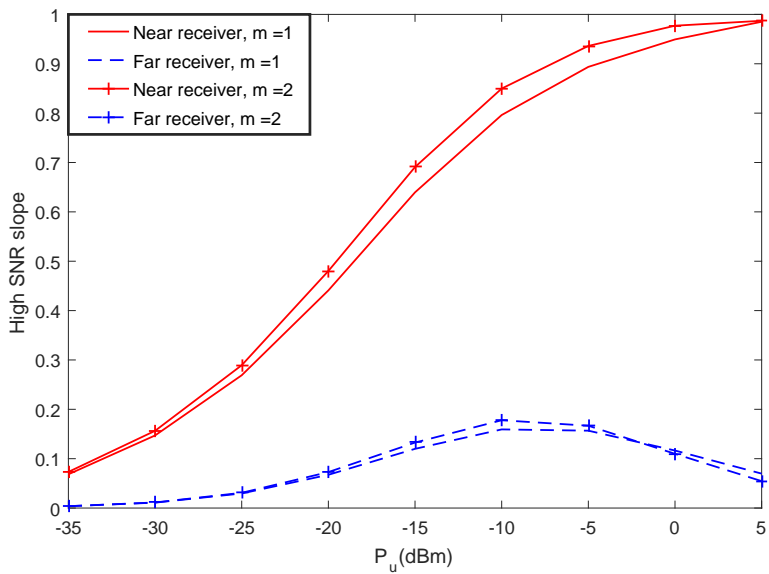}
\caption{High SNR slope of NOMA enhanced U2X framework versus transmit power in both NLoS and LoS scenarios, where the fading parameters $m=1, 2$.}
\label{high SNR slope_fig 7}
\end{figure*}

\section{Conclusions}

In this article, the application of NOMA enhanced U2X frameworks was proposed. Specifically, stochastic geometry tools were invoked for modeling the space randomness of Rxs. Additionally, new closed-form expressions in terms of outage probability and ergodic rate were derived for characterizing the performance in NOMA enhanced U2X frameworks. Diversity orders and high SNR slopes were obtained to evaluate the system performance. The performance of OMA enhanced U2X frameworks were also derived as the benchmark schemes. It was analytically demonstrated that the NOMA enhanced U2X frameworks is capable of outperforming OMA enhanced U2X frameworks. An important future direction is to add the 3-D distribution of interference sources to include other interfering U2X clusters by Poisson hard core process.

\numberwithin{equation}{section}
\section*{Appendix~A: Proof of Theorem~\ref{Theorem1:Outage far UAV stochastic}} \label{Appendix:As}
\renewcommand{\theequation}{A.\arabic{equation}}
\setcounter{equation}{0}

First, the outage probability of the far Rx can be written as follows:
\begin{equation}\label{appendixA1_outage_defination}
{\rm{P}}_v = \mathbb{E} \left( {\frac{{{P_u}{{\left| {{h_v}} \right|}^2}d_v^{ - \alpha }\alpha _v^2}}{{{\sigma ^2} + {P_u}{{\left| {{h_v}} \right|}^2}d_v^{ - \alpha }\alpha _w^2}} < {\varepsilon _v}} \right).
\end{equation}

After some algebraic manipulations, the above outage probability can be rewritten to
\begin{equation}\label{appendixA2_outage_2}
{\rm{P}}_v =  \mathbb{E} \left( {{{\left| {{h_v}} \right|}^2} < \frac{{{\varepsilon _v}{\sigma ^2}d_v^{ \alpha }}}{{{P_u}(\alpha _v^2 - {\varepsilon _v}\alpha _w^2)}}} \right),
\end{equation}
if $\alpha _w^2 - \alpha _w^2{\varepsilon _v} > 0$ holds, otherwise ${\rm{P}}_v =1$.

Recall that the far Rx is located according to a HPPP in the large hollow sphere $\mathcal{V}_2^3$ within the radius $R$ and $D$, and the small-scale fading follows Nakagami-$m$ distribution, the outage probability can be transformed into
\begin{equation}\label{appendixA3_outage_expression}
\begin{aligned}
{\rm{P}}_v &= {\rm{1}} - \frac{3}{{4\pi ({D^3} - {R^3})}}\int\limits_R^D {} \int\limits_0^{\pi } {} \int\limits_0^{2\pi } {{r^2}} \sum\limits_{n = 0}^{m - 1} {\frac{{(m{M_v}{\sigma ^2}r^{ \alpha })^n}}{{n!}}} \\
&\times \exp \left( { - m{M_v}{\sigma ^2}r^{ \alpha }} \right) \sin (\varphi)  d\theta d\varphi dr,
\end{aligned}
\end{equation}
where $\theta$ and $\varphi$ denote the horizontal angle and vertical angle between the Rx and the Tx-UAV.

After some algebraic handling, the outage can be further transformed into
\begin{equation}\label{appendixA4_outage_exact expression}
\begin{aligned}
{\rm{P}}_v &= {\rm{1}} - \frac{3}{{({D^3} - {R^3})}}\sum\limits_{n = 0}^{m - 1} {\frac{{{{(m{M_v}{\sigma ^2})}^n}}}{{n!}}} \\
& \times \int\limits_R^D {} {r^{\alpha n + 2}}\exp \left( { - m{M_v}{\sigma ^2}{r^\alpha }} \right)dr.
\end{aligned}
\end{equation}

By substituting $t = m{M_v}{\sigma ^2}{r^\alpha }$ into \eqref{appendixA4_outage_exact expression}, the outage probability can be simplified as follows:
\begin{equation}\label{appendixA5_outage_closed expression}
{\rm{P}}_v^O = 1 - \frac{{3{{(m{M_v}{\sigma ^2})}^{ - \frac{3}{\alpha }}}}}{{\alpha ({D^3} - {R^3})}}\sum\limits_{n = 0}^{m - 1} {\frac{1}{{n!}}} \int\limits_{m{M_v}{\sigma ^2}{R^\alpha }}^{m{M_v}{\sigma ^2}{D^\alpha }} {} {t^{n + \frac{3}{\alpha } - 1}}\exp \left( { - t} \right)dr,
 \end{equation}
and then the closed-form expression in \eqref{outage analytical results far in theorem1} can be obtained.

\numberwithin{equation}{section}
\section*{Appendix~B: Proof of Corollary~\ref{corollary1:Outage far UAV stochastic}} \label{Appendix:Bs}
\renewcommand{\theequation}{B.\arabic{equation}}
\setcounter{equation}{0}

The asymptotic result of the far Rx $v$ is worth estimating. In the asymptotic outage probability, the transmit SNR between the Tx-UAV and Rxs obeys ${\frac{P_u}{\sigma^2}  \to \infty }$. Recall that $\mathop {\lim }\limits_{x \to 0 } \left( {1 - {e^{ - x}}} \right) \approx x$, and the outage probability of the far Rx can be approximated at the high transmit SNR regime as follows:

\begin{equation}\label{Appendix B1 outage asymptotic}
\begin{aligned}
{\rm{{\hat P}}}_v &= {\rm{1}} - \frac{3}{{({D^3} - {R^3})}}\sum\limits_{n = 0}^{m - 1} {\frac{{{{(m{M_v}{\sigma ^2})}^n}}}{{n!}}} \\
&\times \int\limits_R^D {} {r^{\alpha n + 2}}(1 - m{M_v}{\sigma ^2}{r^\alpha })dr \\
&= 1 - \frac{3}{{({D^3} - {R^3})}}\sum\limits_{n = 0}^{m - 1} {\frac{{{{(m{M_v}{\sigma ^2})}^n}}}{{n!}}\frac{{{D^{\alpha n + 3}} - {R^{\alpha n + 3}}}}{{\alpha n + 3}}} \\
& + \frac{3}{{({D^3} - {R^3})}}\sum\limits_{n = 0}^{m - 1} {\frac{{{{(m{M_v}{\sigma ^2})}^{n + 1}}}}{{n!}}\frac{{{D^{\alpha (n + 1) + 3}} - {R^{\alpha (n + 1) + 3}}}}{{\alpha (n + 1) + 3}}},
\end{aligned}
\end{equation}
and the corollary is proved.

\numberwithin{equation}{section}
\section*{Appendix~C: Proof of Corollary~\ref{corollary: Outage far UAV stochastic m infinity}} \label{Appendix:Cs}
\renewcommand{\theequation}{C.\arabic{equation}}
\setcounter{equation}{0}

Applying limits $m \to \infty$, one can know that ${\left| {{h_v}} \right|^2} = 1$ and ${\left| {{h_w}} \right|^2} = 1$, and thus the outage probability is only affected on the distance of the paired NOMA Rxs. Therefore, we can have the outage probability of the far Rx conditioned on the distance as follows:
\begin{equation}\label{asympto w in corollary m infinity}
\begin{aligned}
{\rm{P}}_v^\infty (d) = \left\{ \begin{array}{l}
1,\,\,d > z_f,\\
\frac{1}{2},\,\,d = z_f,\\
0,\,\,d < z_f,
\end{array} \right.
\end{aligned}
\end{equation}
where $z_f = {\left( {\frac{{{P_u}(\alpha _v^2 - {\varepsilon _v}\alpha _w^2)}}{{{\varepsilon _v}{\sigma ^2}}}} \right)^{\frac{1}{\alpha }}}$.

Recall that far Rxs are located in the hollow space between the radius from $R$ to $D$, and applying the threshold $z$ to the distance distribution, the outage probability of the far Rx can be written to

\begin{equation}\label{theoretical in appendix overall}
{\rm{P}}_v^\infty  = \left\{ \begin{array}{l}
1,\,\,{z_f} < R,\\
\int\limits_R^{{z_f}} {} \frac{3}{{({D^3} - {R^3})}}{r^2}dr,\,\,R < {z_f} < D,\\
0,\,\,D < {z_f}.
\end{array} \right.
\end{equation}
After some algebraic manipulations, the result can be transformed into
\begin{equation}\label{minf final result in appendix overall}
{\rm{P}}_v^\infty  = \left\{ \begin{array}{l}
1,\,\,{z_f} < R,\\
\frac{{z_f^3 - {R^3}}}{{({D^3} - {R^3})}},\,\,R < {z_f} < D,\\
0,\,\,D < {z_f},
\end{array} \right.
\end{equation}
and thus, the corollary is proved.

\numberwithin{equation}{section}
\section*{Appendix~D: Proof of Corollary~\ref{corollary3:ergodic rate near UAV stochastic}} \label{Appendix:Ds}
\renewcommand{\theequation}{D.\arabic{equation}}
\setcounter{equation}{0}

The proof start by providing the ergodic rate of the near Rx $w$ as follows:
\begin{equation}\label{w ergodic exp}
\begin{aligned}
&{R_w} = {\mathbb E}\left\{ {{{\log }_2}\left( {1 + SIN{R_w}\left( {{x_w}} \right)} \right)} \right\} \\
&=  - \int\limits_0^\infty  {{{\log }_2}(1 + {x_{w}})} d\left( {1 - {\rm F}\left( {{x_{w}}} \right)} \right)\\
& = \frac{1}{{\ln \left( 2 \right)}}\int\limits_0^\infty  {\frac{{1 - {\rm F}\left( {{x_{w}}} \right)}}{{1 + {x_{w}}}}} d{x_{w}}.
\end{aligned}
\end{equation}

The cumulative distribution function (CDF) of the near Rx $w$ can be calculated as
\begin{equation}\label{CDF of w}
\begin{aligned}
{F_{{x_w}}}\left( {{x_w}} \right)& = \frac{3}{{({R^3} - r_0^3)}}\sum\limits_{n = 0}^{m - 1} {\frac{{{{\left( {\frac{{mx{\sigma ^2}}}{{{P_u}\alpha _w^2}}} \right)}^n}}}{{n!}}} \\
&\times \int\limits_{{r_0}}^R {} {r^{\alpha n + 2}}\exp \left( { - \frac{{mx{\sigma ^2}}}{{{P_u}\alpha _w^2}}{r^\alpha }} \right)dr,
\end{aligned}
\end{equation}

By substituting \eqref{CDF of w} into \eqref{w ergodic exp}, and recall that the integral $\int\limits_0^\infty  {\frac{\rho }{{1 + x}}dx}$ does not exist, where $\rho$ is a constant.
The achievable ergodic rate can be illustrated as
\begin{equation}\label{expression ergodic rate}
{R_w} = \frac{1}{{\ln \left( 2 \right)}}\int\limits_0^\infty  {\frac{{\frac{3}{{{R^3} - r_0^3}}\sum\limits_{n = 0}^{m - 1} {\frac{{{{\left( {Cx} \right)}^n}}}{{n!}}} \int_{{r_0}}^R {{r^{\alpha n + 2}}\exp \left( { - Cx{r^\alpha }} \right)} dr}}{{1 + {x}}}} d{x},
\end{equation}
where $C = \frac{{m{\sigma ^2}}}{{{P_u}\alpha _w^2}}$.
By substituting $t = Cx{r^\alpha }$ into \eqref{expression ergodic rate}, and after some algebraic manipulations, the ergodic rate can be transformed into

\begin{equation}\label{full express}
\begin{aligned}
{R_w}& = \frac{3}{{\alpha \ln \left( 2 \right)\left( {{R^3} - r_0^3} \right)}}\int\limits_0^\infty  {} \sum\limits_{n = 0}^{m - 1} {\frac{{{{\left( {Cx} \right)}^{ - \frac{3}{\alpha }}}}}{{n!}}} \\
&\times  \frac{{\left( {{\rm{\gamma }}\left( {n + \frac{3}{\alpha }+1,C{R^\alpha }x} \right) - {\rm{\gamma }}\left( {n + \frac{3}{\alpha }+1,Cr_0^\alpha x} \right)} \right)}}{{1 + x}}dx.
\end{aligned}
\end{equation}

One can know from \eqref{full express} that the exact expression of CDF includes lower incomplete Gamma function, which is challenging to calculate. Thus, by utilizing the exponential series
expansion, the lower incomplete Gamma function ban be expanded as

\begin{equation}\label{incomplete gamma trans}
\begin{aligned}
{\rm{\gamma }}\left( {n + \frac{3}{\alpha }+1 ,C{R^\alpha }x} \right) &= \sum\limits_{k = 0}^\infty  {\frac{{\Gamma \left( {n + \frac{3}{\alpha }} \right)}}{{\Gamma \left( {n + \frac{3}{\alpha } + 1 + k} \right)}}} \\
&\times {\left( {C{R^\alpha }x} \right)^{n + \frac{3}{\alpha } + k}}\exp ( - C{R^\alpha }x).
\end{aligned}
\end{equation}

After some algebraic handling, the achievable ergodic rate can be transformed into

\begin{equation}\label{ergodic rate before last}
\begin{aligned}
{R_w} &= \underbrace {\frac{3}{{\alpha \ln \left( 2 \right)\left( {{R^3} - r_0^3} \right)}}\sum\limits_{n = 0}^{m - 1} {\frac{1}{{n!}}} \sum\limits_{k = 0}^\infty  {\frac{{\Gamma \left( {n + \frac{3}{\alpha }} \right)}}{{\Gamma \left( {n + \frac{3}{\alpha } + 1 + k} \right)}}  {C^{n + k}}{R^{\alpha n + 3 + \alpha k}}} \int\limits_0^\infty  {\frac{{{x^{n + k}}\exp ( - C{R^\alpha }x)}}{{1 + x}}} dx}_{{J_1}} \\
& - \underbrace {\frac{3}{{\alpha \ln \left( 2 \right)\left( {{R^3} - r_0^3} \right)}}\sum\limits_{n = 0}^{m - 1} {\frac{1}{{n!}}} \sum\limits_{k = 0}^\infty  {\frac{{\Gamma \left( {n + \frac{3}{\alpha }} \right)}}{{\Gamma \left( {n + \frac{3}{\alpha } + 1 + k} \right)}} {C^{n + k}}r_0^{\alpha n + 3 + \alpha k}} \int\limits_0^\infty  {\frac{{{x^{n + k}}\exp ( - Cr_0^\alpha x)}}{{1 + x}}} dx}_{{J_2}}.
\end{aligned}
\end{equation}

Based on \cite[eq. (3.352.4)]{Table_of_integrals} and applying polynomial expansion manipulations, $J_1$ and $J_2$ can be expressed as
\begin{equation}\label{J1}
\begin{aligned}
{J_1} =&  \frac{3}{{\alpha \ln \left( 2 \right)\left( {{R^3} - r_0^3} \right)}}\sum\limits_{n = 0}^{m - 1} {\frac{1}{{n!}}} \sum\limits_{k = 0}^\infty  {\frac{{\Gamma \left( {n + \frac{3}{\alpha }} \right)\Gamma \left( {n + 1 + k} \right)}}{{\Gamma \left( {n + \frac{3}{\alpha } + 1 + k} \right)}}} \\
&\times {C^{n + k}}{R^{\alpha n + 3 + \alpha k}} \exp (C{R^\alpha })\Gamma \left( { - k - n,C{R^\alpha }} \right),
\end{aligned}
\end{equation}
and
\begin{equation}\label{J2_appendixC}
\begin{aligned}
{J_2} =& \frac{3}{{\alpha \ln \left( 2 \right)\left( {{R^3} - r_0^3} \right)}}\sum\limits_{n = 0}^{m - 1} {\frac{1}{{n!}}} \sum\limits_{k = 0}^\infty  {\frac{{\Gamma \left( {n + \frac{3}{\alpha }} \right)\Gamma \left( {n + 1 + k} \right)}}{{\Gamma \left( {n + \frac{3}{\alpha } + 1 + k} \right)}}}\\
& \times {C^{n + k}}r_0^{\alpha n + 3 + \alpha k} \exp (Cr_0^\alpha )\Gamma \left( { - k - n,Cr_0^\alpha } \right),
\end{aligned}
\end{equation}
where $\Gamma \left( { ,} \right)$ denotes upper incomplete Gamma function, and the proof of corollary is completed.

\bibliographystyle{IEEEtran}%
\bibliography{IEEEabrv,bib2018}

\end{document}